\documentclass[12pt,preprint]{aastex}

\def\kms{\ifmmode{\rm km\th s^{-1}}\else km\th s$^{-1}$\fi}
\def\th{\thinspace}

\shortauthors{Boden et al.}
\shorttitle{HD~9939}

\begin{document}

\title{A Physical Orbit for the High Proper Motion Binary HD~9939}

%%\email{**Draft \today**}

\author{Andrew F.~Boden\altaffilmark{1,2},
        Guillermo Torres\altaffilmark{3},
	David W.~Latham\altaffilmark{3}
}

\altaffiltext{1}{Michelson Science Center, California
Institute of Technology, 770 South Wilson Ave., Pasadena CA 91125}

\altaffiltext{2}{Department of Physics and Astronomy, Georgia State
University, 29 Peachtree Center Ave., Science Annex, Suite 400,
Atlanta GA 30303}

\altaffiltext{3}{Harvard-Smithsonian Center for Astrophysics, 60
Garden St., Cambridge MA 02138}

%%\altaffiltext{4}{Present address: Palomar Observatory, California
%%Institute of Technology, 770 South Wilson Ave., Pasadena CA 91125}

\email{bode@ipac.caltech.edu}

\begin{abstract}

We report spectroscopic and interferometric observations of the
high-proper motion double-lined binary system HD~9939, with an orbital
period of approximately 25~days.  By combining our radial-velocity and
visibility measurements we estimate the system physical orbit and
derive dynamical masses for the components of $M_A = 1.072 \pm
0.014$~M$_{\sun}$ and $M_B = 0.8383 \pm 0.0081$~M$_{\sun}$; fractional
errors of 1.3\% and 1.0\%, respectively.  We also determine a system
distance of $42.23 \pm 0.21$~pc, corresponding to an orbital parallax
of $\pi_{\rm orb} = 23.68 \pm 0.12$~mas.  The system distance and the
estimated brightness difference between the stars in $V$, $H$, and $K$
yield component absolute magnitudes in these bands.  By spectroscopic
analysis and spectral energy distribution modeling we also estimate
the component effective temperatures and luminosities as $T_{\rm
eff}^A = 5050 \pm 100$~K and $T_{\rm eff}^B = 4950 \pm 200$~K and
$L_A$ = 2.451 $\pm$ 0.041 $L_{\sun}$ and $L_B$ = 0.424 $\pm$ 0.023
$L_{\sun}$.

Both our spectral analysis and comparison with stellar models suggest
that HD~9939 has elemental abundances near solar values.  Further,
comparison with stellar models suggests the HD~9939 primary has
evolved off the main sequence and appears to be traversing the
Hertzsprung gap as it approaches the red giant phase of its evolution.
Our measurements of the primary properties provide new empirical
constraints on stellar models during this particularly dynamic
evolutionary phase.  That HD~9939 is currently in a relatively
short-lived evolutionary state allows us to estimate the system age as
9.12 $\pm$ 0.25 Gyr. In turn the age and abundance of the system place
a potentially interesting, if anecdotal, constraint on star formation
in the galactic disk.

\end{abstract}


\section{Introduction}
\label{sec:introduction}

Accurate determinations of the physical properties of stars in binary
systems (mass, radius, luminosity, elemental abundance, etc.) provide
for fundamental tests of models of stellar structure and stellar
evolution.  The most basic of those stellar properties is the mass,
typically available only through analysis of dynamical interactions in
general and binary systems in particular.  Several dozen eclipsing
binary systems have component mass and radius determinations that are
accurate to 1--2\% \citep[e.g.,][]{Andersen1991}, and show that
main-sequence models for stars with masses in the range from about
1~M$_{\sun}$ to 10~M$_{\sun}$ and elemental abundances near solar are
in fairly good agreement with the observations.  However, models for
elemental compositions that are much different from solar are largely
untested by observations due to a lack of suitable systems or a lack
of accurate measurements.  Consequently in 2000 we initiated a program
to measure the properties of metal-poor stars in binary systems.  The
first results from that program were reported in \citet[hereafter
Paper~1]{Torres2002}.  Here we follow Paper~1 with a report on our
measurements of the high-proper motion binary system HD~9939.

\objectname[HD 9939]{HD~9939} (aka G34-39, HIP 7564, SAO 74830) is a
relatively short-period (25 d) binary system.  HD~9939 was identified
as a high proper-motion object in the Luyten \citep[][and references
therein]{Luyten1980,Salim2003} and Lowell
\citep{Giclas1971,Giclas1978} proper motion surveys.  HD~9939 was
included as part of the high-proper motion radial velocity survey
program of Carney \& Latham et al \citep[see][]{Carney1987}, where it
was identified as a single \citep{Latham1988} and eventually a
double-lined spectroscopic binary \citep[][and references
therein]{Goldberg2002}.  (Separately HD~9939 was identified as a
suspected binary system in the Mt.~Wilson survey of Fouts and Sandage
\citep[][and references therein]{Fouts1987}).  The consensus spectral
classification for the HD~9939 system is K0 IV by \citet[][based on
observations by Kuiper]{Bidelman1985} and \citet[][responsible for the
luminosity classification]{Yoss1961}.  Analysis of the spectra taken
in the above-mentioned Carney \& Latham project yielded an estimated
HD~9939 system metallicity [Fe/H] = -0.51 dex (hence its inclusion in
our program), and an estimated effective temperature of 4770 K.
However both of these estimates were flagged as highly uncertain
\citep{Carney1994}.  Based on the double-lined orbit of
\citet[][hereafter G2002]{Goldberg2002}, \citet{Mazeh2003} included
HD~9939 in a statistical study of mass ratios in binary systems.

HD~9939 was identified as a candidate for our program on metal-poor
systems in 2000, and observed with the Palomar Testbed Interferometer
(PTI) and Harvard-Smithsonian Center for Astrophysics (CfA)
telescopes.  Herein we report the physical orbit estimation for
HD~9939 based on these observations, and the analysis of the system
components based on the physical orbit.

\section{Observations}
\label{sec:observations}

\subsection{Spectroscopy}
\label{sec:spectroscopy}

HD~9939 was observed at the CfA with an echelle spectrograph on the
1.5-m Wyeth reflector at the Oak Ridge Observatory (eastern
Massachusetts), and occasionally also with a nearly identical
instrument on the 1.5-m Tillinghast reflector at the F.\ L.\ Whipple
Observatory (Arizona).  A single echelle order was recorded using
intensified photon-counting Reticon detectors, spanning about 45\,\AA\
at a central wavelength of 5188.5\,\AA, which includes the
\ion{Mg}{1}~b triplet. The resolving power of these instruments is
$\lambda/\Delta\lambda \approx 35,\!000$, and the signal-to-noise
(S/N) ratios achieved range from 11 to about 47 per resolution element
of 8.5~km~s$^{-1}$.  A total of 62 spectra were obtained between 6
September 1982 and 4 February 2002, spanning 19.4 years and 281 system
periods, with many spectra taken relatively recently as HD~9939 was
included in our joint program.

Radial velocities (RVs) were determined from the spectra with the
two-dimensional cross-correlation algorithm TODCOR \citep{ZM94}, which
uses two templates, matched to each component of the binary. The
templates were selected from an extensive library of calculated
spectra based on model atmospheres by R.\ L.\
Kurucz\footnote{Available at {\tt http://cfaku5.cfa.harvard.edu}.},
computed for us by Jon Morse \citep[see also][]{Nordstrom1994,
Latham2002}. These calculated spectra are available for a wide range
of effective temperatures ($T_{\rm eff}$), rotational velocities
($v_{\rm rot}$), surface gravities ($\log g$), and metallicities. The
optimum template for each star was determined from grids of
cross-correlations over broad ranges in temperature and rotational
velocity (since these are the parameters that affect the radial
velocities the most), seeking to maximize the average correlation
weighted by the strength of each exposure. Surface gravities of $\log
g = 4.0$ for the primary and $\log g = 4.5$ for the secondary were
adopted from the results of \S\ref{sec:physics}.  The grids of
correlations were repeated for a range of metallicities, and we found
that our spectra have a strong preference for a composition near
solar, as opposed to the more negative but uncertain value of ${\rm
[m/H]} = -0.51$ reported by \cite{Carney1994}. Consequently we adopted
the solar value for determining the RVs.  The temperature estimates
from our spectra for this metallicity are $T^A_{\rm eff} = 5050 \pm 100$~K
and $T^B_{\rm eff} = 4950 \pm 200$~K for the primary and secondary,
respectively. The rotational broadening of the stars was found to be
negligible.  The light ratio inferred from these spectra at the mean
wavelength of our observations is $\ell_B/\ell_A = 0.16 \pm 0.03$.

Following our procedures in Paper~1 we tested our velocities for
systematic effects by means of numerical simulations. The corrections
were found to be small (typically well under 1~km~s$^{-1}$), but were
applied nevertheless for consistency. The final velocities including
these corrections are presented in Table~\ref{tab:RVdata}. These
measurements, and the orbital solutions derived from them below,
supersede those obtained by G2002 %%\cite{Goldberg2002}
that were based on a subset of the present spectra.

\begin{table}
\dummytable\label{tab:RVdata}
\end{table}

\subsection{Interferometry}
\label{sec:interferometry}
As in Paper~I, the interferometric observable used for these
measurements is the fringe contrast or {\em visibility} (squared) of
an observed brightness distribution on the sky.  PTI was used to make
the interferometric measurements presented here; PTI is a
long-baseline $H$ (1.6$\mu$m) and $K$-band (2.2$\mu$m) interferometer
located at Palomar Observatory, and described in detail elsewhere
\citep{Colavita1999a}.  The analysis of such data on a binary system
is discussed in detail in previous work \citep[e.g.][]{Boden2000} and
elsewhere \citep[e.g.][]{Hummel1998} and is not repeated here.

HD~9939 was observed in conjunction with objects in our calibrator
list by PTI in $K$ band ($\lambda \sim 2.2 \mu$m) on 91 nights between
9 August 2000 and 13 December 2004, a dataset spanning roughly 4.3
years and 62 orbital periods.  Additionally, HD~9939 was observed by
PTI in $H$ band ($\lambda \sim 1.6 \mu$m) on six nights between 15
September 2000 and 18 August 2002, spanning roughly two years and 28
system periods.  On each night HD~9939 and calibration objects were
typically observed repeatedly, and each observation, or scan, was
approximately 130 sec long.  For each scan we computed a mean $V^2$
value from the scan data, and the error in the $V^2$ estimate from the
rms internal scatter \citep{Colavita1999b,Boden1998}.  HD~9939 was
observed in combination with one or more calibration sources within
$\sim$ 10$^{\circ}$ on the sky.  Here we have used two stars as
calibration objects: \objectname[HD 7034]{HD~7034} (putative F0 V),
\objectname[HD 7964]{HD~7964} (putative A3 V); Table
\ref{tab:calibrators} lists the relevant physical parameters for the
calibration objects.  Calibrating our interferometric data using the
these objects results in 542 $K$ and 60 $H$ visibility scans on
HD~9939.  The calibrated $V^2$ data on HD~9939 are summarized in
Table~\ref{tab:V2Table}.

\begin{deluxetable}{ccccc}
\tabletypesize{\small}
\tablecolumns{5}
\tablewidth{0pc}

\tablecaption{PTI HD~9939 $V^2$ Calibration Objects in our
Analysis.  The relevant parameters for our two calibration objects are
summarized.  The apparent diameter values are determined from spectral
energy distribution modeling based on archival broad-band photometry,
spectral energy distribution templates from \cite{Pickles1998}, and
visibility measurements with PTI.
\label{tab:calibrators}
}

\tablehead{
\colhead{Object} & \colhead{Spectral} & \colhead{Star}      & \colhead{Separation}  & \colhead{Adopted Model} \\
\colhead{Name}   & \colhead{Type}     & \colhead{Magnitude} & \colhead{from HD~9939}& \colhead{Diameter (mas)}
}

\startdata
HD 7034   & F0 V     & 5.2 V/4.4 K & 8.5$^{\circ}$  & 0.51 $\pm$ 0.02   \\
HD 7964   & A3 V     & 4.7 V/4.5 K & 4.5$^{\circ}$  & 0.42 $\pm$ 0.04   \\
\enddata

\end{deluxetable}

\begin{table}
\dummytable\label{tab:V2Table}
\end{table}

\section{Orbit Determination}
\label{sec:orbit}

As in Paper~1, we have estimated the HD~9939 orbit based separately on
the CfA spectroscopy, the PTI visibilities, and jointly with these two
datasets.

\begin{figure}
\epsscale{0.8}
%%\plotone{figures/hd9939.trace.eps}
\plotone{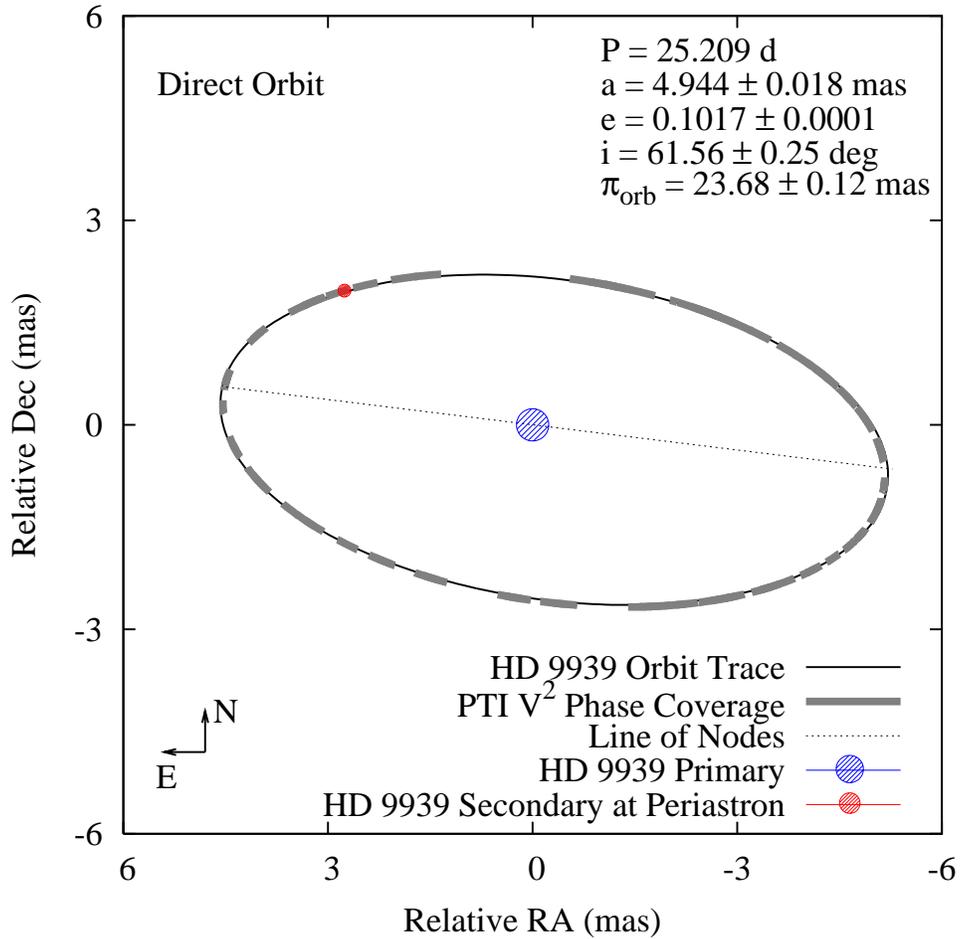}
\caption{Visual Orbit of HD~9939.  The relative visual orbit model of
HD~9939 is shown, with the primary and secondary objects rendered at
T$_0$ (periastron).  The heavy lines along the relative orbit indicate
areas where we have phase coverage in our K-band PTI data (they are
not separation vector estimates); our data sample essentially all
phases of the orbit well, leading to a reliable orbit determination.
Component diameter values are estimated (see \S\ref{sec:physics}) and
rendered to scale.
\label{fig:hd9939_orbit}}
\end{figure}

Figure \ref{fig:hd9939_orbit} depicts the relative visual orbit of the
HD~9939 system, with the primary component rendered at the origin, and
the secondary component rendered at periastron.  We have indicated the
phase coverage of our $V^2$ data on the relative orbit with heavy
lines; our data sample essentially all phases of the orbit, leading to
a reliable orbit determination.  The size of the HD~9939 components
are estimated (see discussion in \S\ref{sec:physics}) and rendered to
scale.

$V^2$ observations are subject to a point-symmetric inversion
degeneracy (a scene and its mirror inverse produce the same $V^2$), so
astrometric orbits determined with such data have a modulo 180$^\circ$
uncertainty in the apparent orientation on the sky and longitude of
the ascending node ($\Omega$) \citep[e.g.][]{Boden2000,Boden2005}.  In
the case of HD~9939 there is a significant brightness ratio between
the components, so we modeled archival Hipparcos intermediate
astrometric data constrained by the high-precision orbital parameters
determined here (Table~\ref{tab:orbit}) to determine the proper
orientation/$\Omega$ value.  The small angular scale orbit is
marginally resolved in the Hipparcos data.  However fixing all
parameters except the photocentric semi-major axis and $\Omega$
indicates the quadrant value of $\Omega$ as given in
Table~\ref{tab:orbit} and as rendered in Figure \ref{fig:hd9939_orbit}
(i.e. with the secondary North-East of the primary at periastron).
Therefore we believe the proper orientation of the HD~9939 orbit is as
rendered, and $\Omega$ is uniquely determined as given in
Table~\ref{tab:orbit}.

\begin{figure}
\epsscale{1.0}
%%\plotone{figures/hd9939.rvPhase.eps}
\plotone{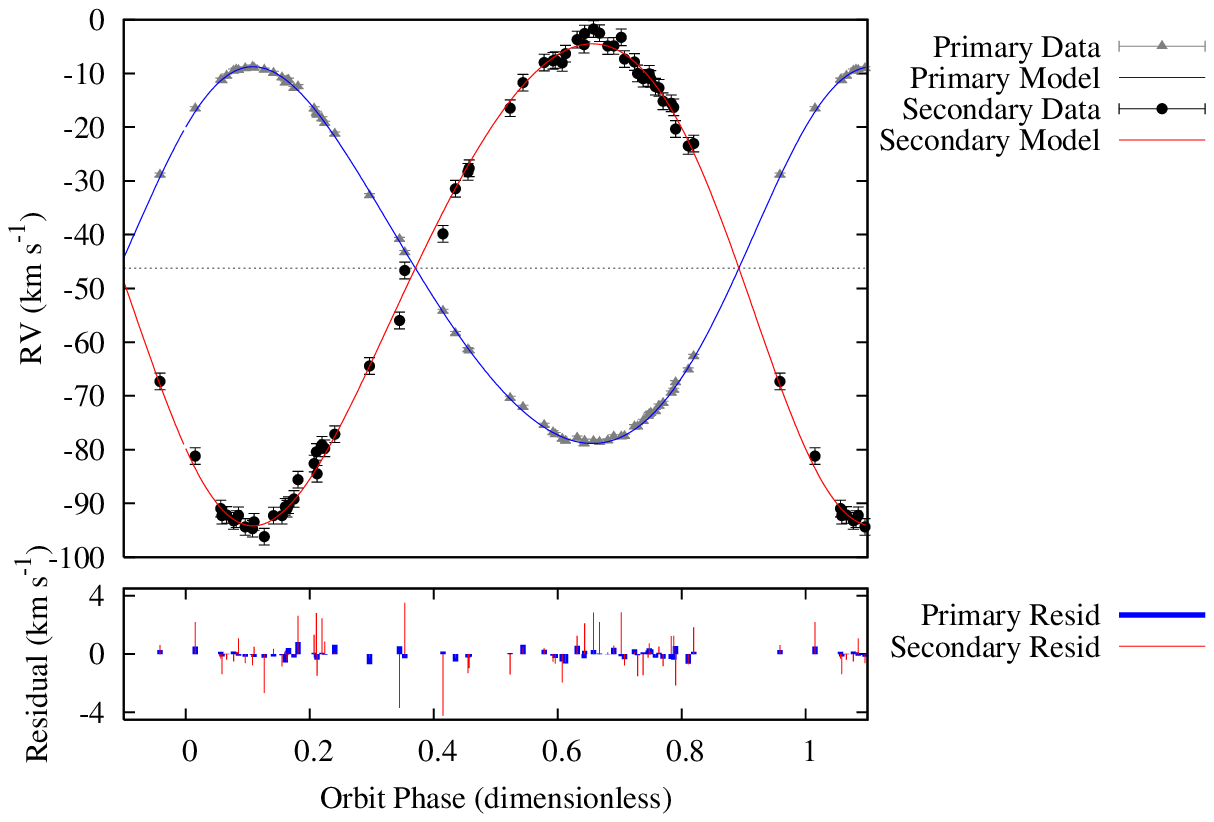}
\caption{Radial Velocity Orbit of HD~9939.  A phase-folded plot of our
radial velocity data and RV predictions from our Full-Fit solution
(Table~\ref{tab:orbit}).  The lower frame gives RV residuals to the
model fit.
\label{fig:hd9939_RVorbit}}
\end{figure}

Figure \ref{fig:hd9939_RVorbit} depicts the phase-folded radial
velocity orbit of HD~9939.  The top frame gives the RV orbit curves
and the measured radial velocities from Table \ref{tab:RVdata}.  The
bottom frame gives the residuals between our RV data and the orbit
model (data - model).  Our solution is in good agreement with the
G2002 double-lined solution derived from many of the same spectra.

\begin{figure}
\epsscale{1.0}
%%\plotone{figures/hd9939.v2Resid.eps}\\
\plotone{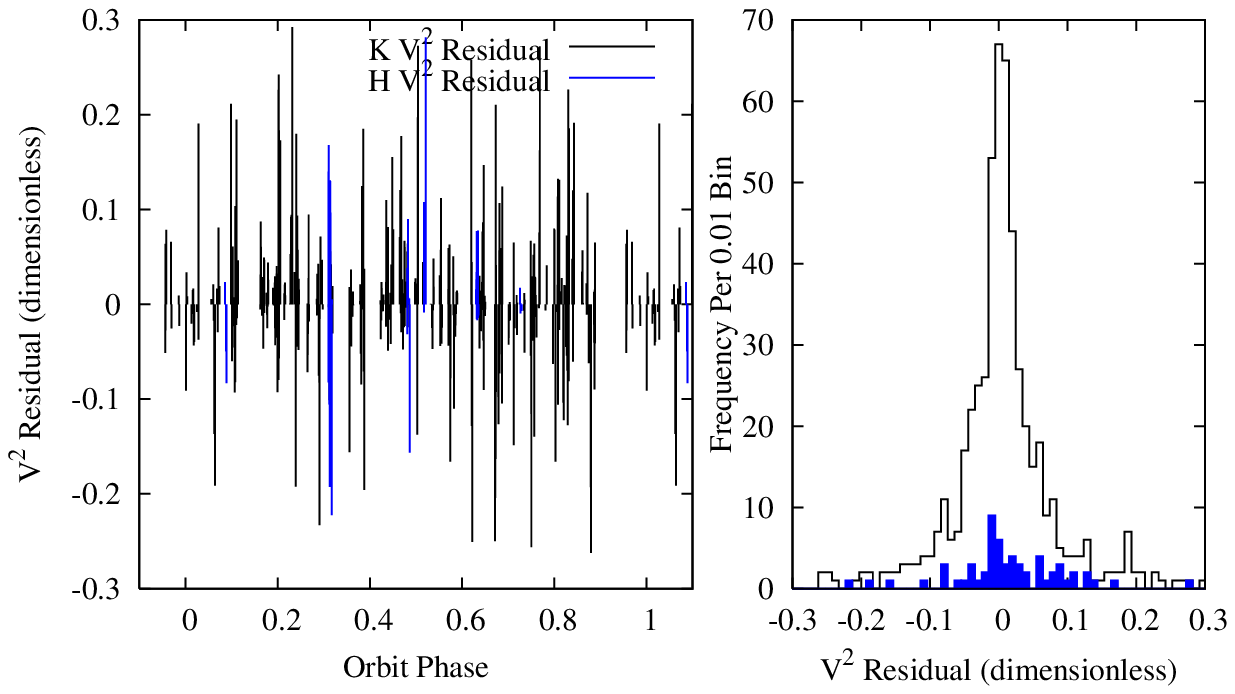}\\
%%\plotone{figures/hd9939.rvResid.eps}
\plotone{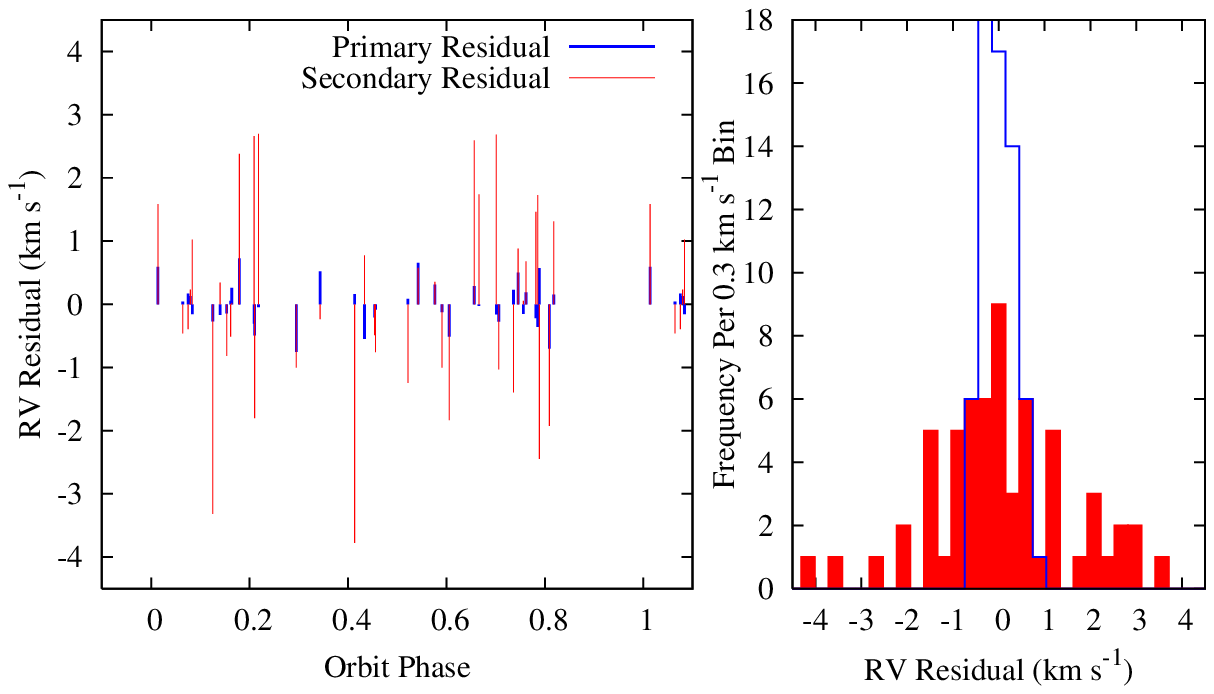}
\caption{Orbit Modeling Residuals for HD~9939.  Phase-folded residual
plots and residual histograms are given for the $V^2$ (top) and RV
(bottom) datasets relative to our Full-Fit solution
(Table~\ref{tab:orbit}).
\label{fig:hd9939Residuals}}
\end{figure}

Table \ref{tab:orbit} summarizes orbit models for HD~9939, including
the model from G2002, and individual and joint fits to our visibility
and radial velocity datasets.  All solutions have component diameters
constrained as noted above.  All orbit parameters are listed with
total one-sigma errors, including contributions from statistical
(measurement uncertainty) and systematic error sources.  In our
analysis the dominant forms of systematic error are: (1) uncertainties
in the calibrator angular diameters (Table \ref{tab:calibrators}); (2)
uncertainty in the center-band operating wavelength ($\lambda_0
\approx$ 2.2 $\mu$m), taken to be 10 nm ($\sim$0.5\%); (3) the
geometrical uncertainty in our interferometric baseline ( $<$ 0.01\%);
and (4) uncertainties in ancillary parameters constrained in our orbit
fitting procedure (i.e.~the angular diameters in all solutions
involving interferometry data).

\begin{deluxetable}{l|c|ccc}
%%%\tabletypesize{\footnotesize}
\tablecolumns{6}
\tablewidth{0pc}
%%\rotate

\tablecaption{Orbital Parameters for HD~9939.  Summarized here are the
apparent orbital parameters for the HD~9939 system as determined by
G2002 and present results.  We give three separate fits to our data:
$K$ V$^2$ only, RV only, and integrated (``Full Fit'').  $\Omega$ is
reported in a Position Angle convention.
\label{tab:orbit}
}

\tablehead{
\colhead{Orbital}        & \colhead{G2002}      & \multicolumn{3}{c}{PTI \& CfA} \\
\colhead{Parameter}      &                      & \colhead{$K$ V$^2$ Only} & \colhead{RV Only}   & \colhead{Full-Fit} }
\startdata
Period (d)               & 25.2125              & 25.20942             & 25.208918             & 25.208958    \\
                         & $\pm$ 0.018          &$\pm$ 4.5 $\times$ 10$^{-4}$ & $\pm$ 8.5 $\times$ 10$^{-5}$ & $\pm$ 7.2 $\times$ 10$^{-5}$ \\
T$_{0}$ (MJD)            & 46039.24             & 51786.63             & 51786.368           & 51786.408 \\
                         & $\pm$ 0.22           &  $\pm$ 0.13          & $\pm$ 0.083         & $\pm$ 0.063 \\
$e$                      & 0.1045               & 0.1006               & 0.1023              & 0.10166 \\
 		         & $\pm$ 0.0046         & $\pm$ 0.0012         & $\pm$ 0.0019        & $\pm$ 0.00097 \\
K$_1$ (km s$^{-1}$)      & 35.17 $\pm$ 0.17     &                      & 34.965 $\pm$ 0.056  & 34.952 $\pm$ 0.055 \\
K$_2$ (km s$^{-1}$)      & 44.89 $\pm$ 0.80     &                      & 44.69 $\pm$ 0.24    & 44.68 $\pm$ 0.24  \\
$\gamma$ (km s$^{-1}$)   & -45.92 $\pm$ 0.12    &                      & -46.148 $\pm$ 0.054 & -46.157 $\pm$ 0.049 \\
$\omega_{1}$ (deg)       & 312.7 $\pm$ 3.0      & 315.8 $\pm$ 1.6      & 312.6 $\pm$ 1.2     & 313.07 $\pm$ 0.88  \\
$\Omega$ (deg)           &                      & 262.55 $\pm$ 0.26    &                     & 262.29 $\pm$ 0.20 \\
$i$ (deg)                &                      & 61.06 $\pm$ 0.31     &                     & 61.56 $\pm$ 0.25 \\
$a$ (mas)                &                      & 4.937 $\pm$ 0.020    &                     & 4.944 $\pm$ 0.018 \\
$\Delta K_{\rm CIT}$ (mag) &                    & 1.983 $\pm$ 0.067    &                     & 1.983 $\pm$ 0.067 \\
$\Delta H_{\rm CIT}$ (mag) &                    &                      &                     & 2.00 $\pm$ 0.14 \\
$\Delta V$ (mag)        &  2.06                 &                      & 1.990 $\pm$ 0.176   & 1.990 $\pm$ 0.176 \\
$\chi^2$/DOF	        &                       & {\em 1.0}            & {\em 1.0}           & {\em 1.0} \\
$\overline{|R_{V^2}|}$/$\sigma_{V^2}$ &         & 0.059/0.11           &                     & 0.057/0.11 \\
$\overline{|R_{RV}|}$/$\sigma_{RV}$ (km s$^{-1}$) &  &                 & 0.70/1.1            & 0.81/1.18 \\
\hline
\enddata

\end{deluxetable}

\section{Physical properties of HD~9939}
\label{sec:physics}

The orbital parameters from Table~\ref{tab:orbit} allow us to compute
many of the physical properties of the HD~9939 system and its
components.  Physical parameters derived from our HD~9939 ``Full-Fit''
integrated visual/spectroscopic orbit and spectral energy distribution
modeling are summarized in Table \ref{tab:physics}.  Notable among
these is the high-precision determination of the component masses for
the system; we estimate the masses of the primary and secondary
components as 1.072 $\pm$ 0.014 and 0.8383 $\pm$ 0.0081 M$_{\sun}$,
respectively.  The relative errors in the primary and secondary
component mass estimates are 1.3\% and 0.97\% respectively.

The Hipparcos catalog lists the parallax of HD~9939 as 23.80 $\pm$
0.86 mas \citep{HIP1997}.  The distance determination to HD~9939 based
on our orbital solution is 42.23 $\pm$ 0.21 pc, corresponding to an
orbital parallax of 23.68 $\pm$ 0.12 mas, consistent with the
Hipparcos result at 0.5\% and 0.1-sigma.

\begin{figure}[p]
\includegraphics[angle=-90,width=14.5cm]{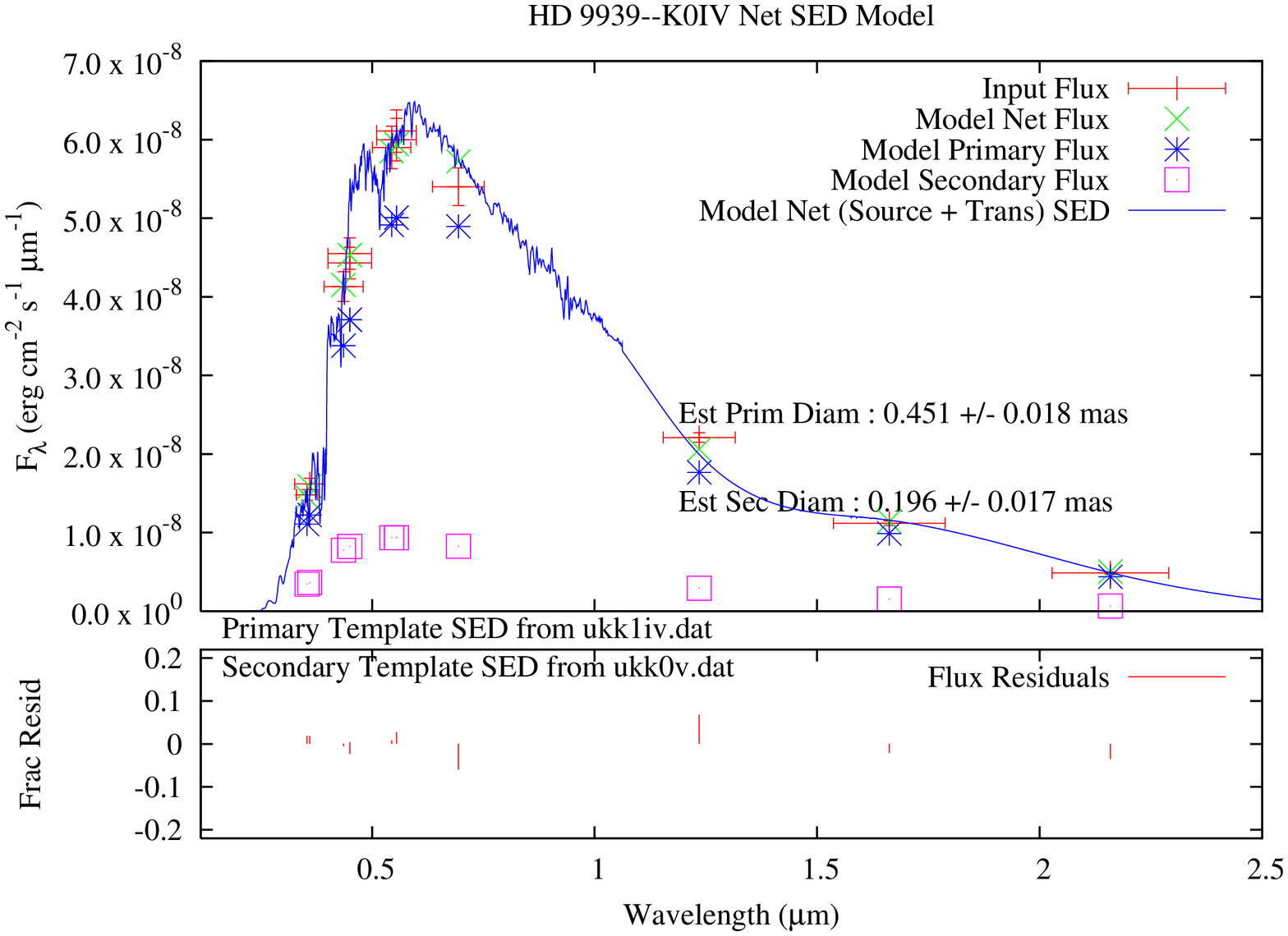}\\
\includegraphics[angle=-90,width=15.5cm]{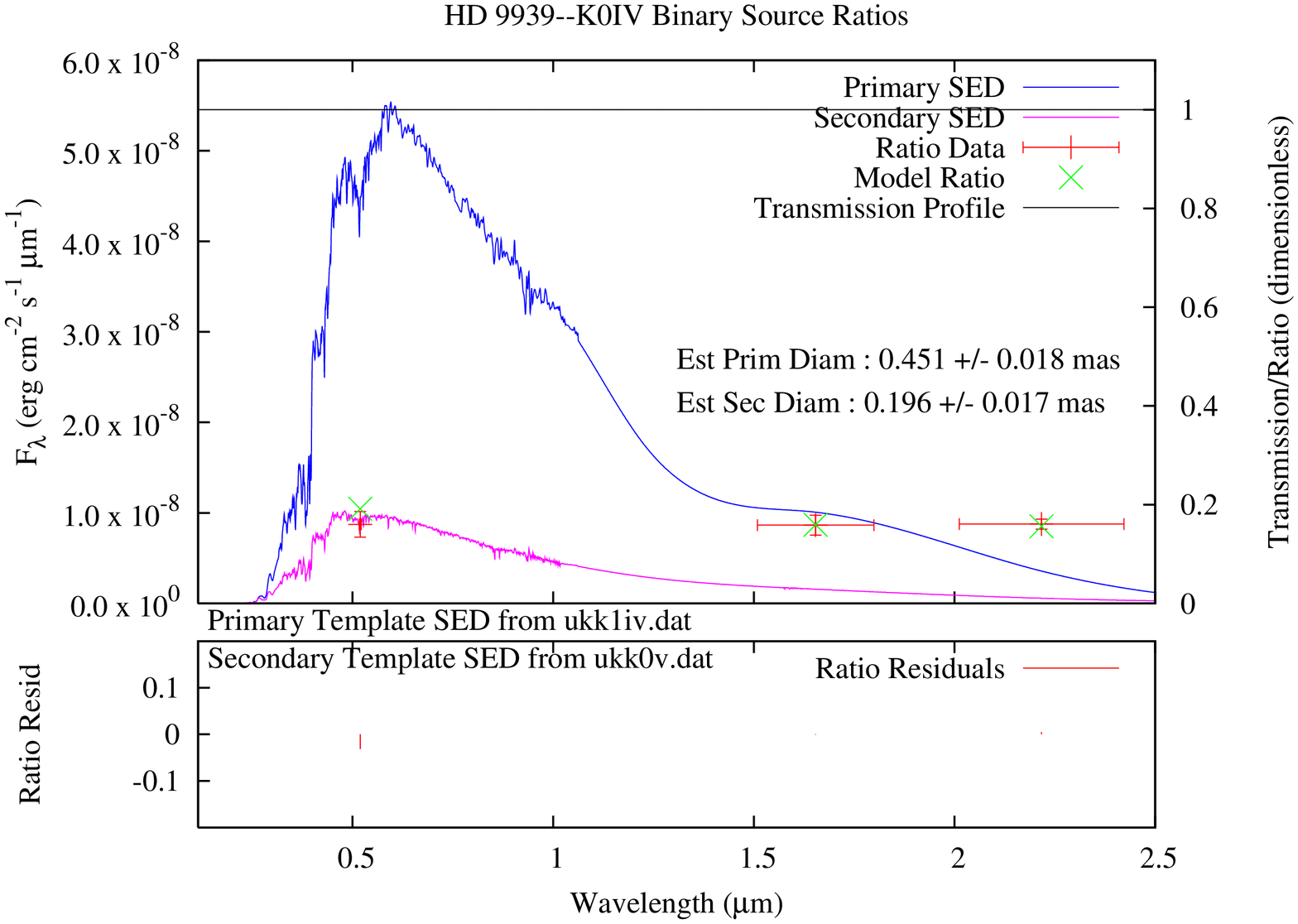}\\
\caption{Spectral Energy Distribution Model for HD~9939.  Here SED
templates from \citet{Pickles1998} have been simultaneously fit to
archival photometry (top) and flux ratio estimates (bottom) from our
visibility and spectroscopic measurements (Table \ref{tab:orbit}).
\label{fig:hd9939SED}}
\end{figure}

At the distance of HD~9939 neither of the system components are
significantly resolved by the PTI $H$ or $K$-band fringe spacings, and
we must resort to spectral energy distribution (SED) modeling to
estimate the component diameters.  As input to the SED modeling we
have used archival photometry and the estimated component flux ratios
in the spectroscopic and visibility data (Table \ref{tab:orbit}).  To
this data we have fit a two-component model with solar abundance SED
templates from \citet{Pickles1998}.  We find a best-fit solution with
Pickles templates matching spectral types K1 IV and K0 V, in good
agreement with the system spectral classification.  This SED model is
depicted in Figure~\ref{fig:hd9939SED}.  Integrating the model
component SEDs we estimate bolometric fluxes of (4.418 $\pm$ 0.059)
$\times$ 10$^{-8}$ and (7.65 $\pm$ 0.40) $\times$ 10$^{-9}$ erg
cm$^{-2}$ s$^{-1}$ for the HD~9939 primary and secondary components
respectively.  When combined with our component model effective
temperatures (\S\ref{sec:spectroscopy}) we estimate apparent angular
diameters of 0.451 $\pm$ 0.018 and 0.196 $\pm$ 0.017 mas for the
primary and secondary components respectively.  At the distance of
HD~9939 the bolometric fluxes imply luminosities of 2.451 $\pm$ 0.041
and 0.424 $\pm$ 0.023 L$_\sun$, and physical radii of 2.048 $\pm$
0.081 and 0.887 $\pm$ 0.071 R$_\sun$ for the primary and secondary
components respectively.

\begin{deluxetable}{ccc}
\tabletypesize{\small} \tablecolumns{3} \tablewidth{0pc}

\tablecaption{Physical Parameters for HD~9939.  Summarized here are
the physical parameters for the HD~9939 system as derived primarily
from the ``Full-Fit'' solution orbital parameters in Table
\ref{tab:orbit}.  Archival system photometry is from
\cite{Mermilliod1994} and \cite{2MASS}.
\label{tab:physics}
}

\tablehead{
\colhead{Physical}   & \colhead{Primary (A)}& \colhead{Secondary (B)} \\
\colhead{Parameter}  & \colhead{Component}  & \colhead{Component}
}

\startdata
a (10$^{-2}$ AU)     & 9.163 $\pm$ 0.026    & 11.712 $\pm$ 0.062  \\
Mass (M$_{\sun}$)    & 1.072 $\pm$ 0.014    & 0.8383 $\pm$ 0.0081   \\
\cline{2-3}
Sp Type              & \multicolumn{2}{c}{K0 IV} \\
System Distance (pc) & \multicolumn{2}{c}{42.23 $\pm$ 0.21} \\
$\pi_{orb}$ (mas)    & \multicolumn{2}{c}{23.68 $\pm$ 0.12} \\
\cline{2-3}
Bolometric Flux (10$^{-9}$ erg cm$^{-2}$ s$^{-1}$) & 
44.18 $\pm$ 0.59 & 7.65 $\pm$ 0.40 \\
T$_{eff}$ (K)        & 5050 $\pm$ 100       & 4950  $\pm$ 200  \\
Model Diameter (mas) & 0.451 $\pm$ 0.018    & 0.196 $\pm$ 0.017  \\
Luminosity (L$_\sun$) & 2.451 $\pm$ 0.041   & 0.424 $\pm$ 0.023  \\
Radius (R$_\sun$)    & 2.048 $\pm$ 0.081    & 0.887 $\pm$ 0.071  \\
$\log$ g             & 3.845 $\pm$ 0.035    & 4.465 $\pm$ 0.070  \\
M$_{K-{\rm CIT}}$ (mag) & 1.896 $\pm$ 0.022 & 3.881 $\pm$ 0.062 \\
M$_{H-{\rm CIT}}$ (mag) & 2.027 $\pm$ 0.023 & 4.027 $\pm$ 0.12  \\
M$_V$ (mag)          & 4.003 $\pm$ 0.033    & 5.99  $\pm$ 0.15  \\
$V$-$K$ (mag)        & 2.080 $\pm$ 0.037    & 2.09 $\pm$ 0.16 \\
\enddata

\end{deluxetable}

\section{Comparison with Stellar Evolution Models}
\label{sec:modelcomp}

With our estimates of the component masses, absolute magnitudes, color
indices, and effective temperatures derived from our measurements and
orbital solution (Table~\ref{tab:physics}), we proceed in this section
to examine the physical state of the HD~9939 components in the context
of recent stellar evolution models from the Yonsei-Yale (Y$^2$)
collaboration \citep{Yi2001,Demarque2004}.

\begin{figure}[p]
\epsscale{0.85}
\plotone{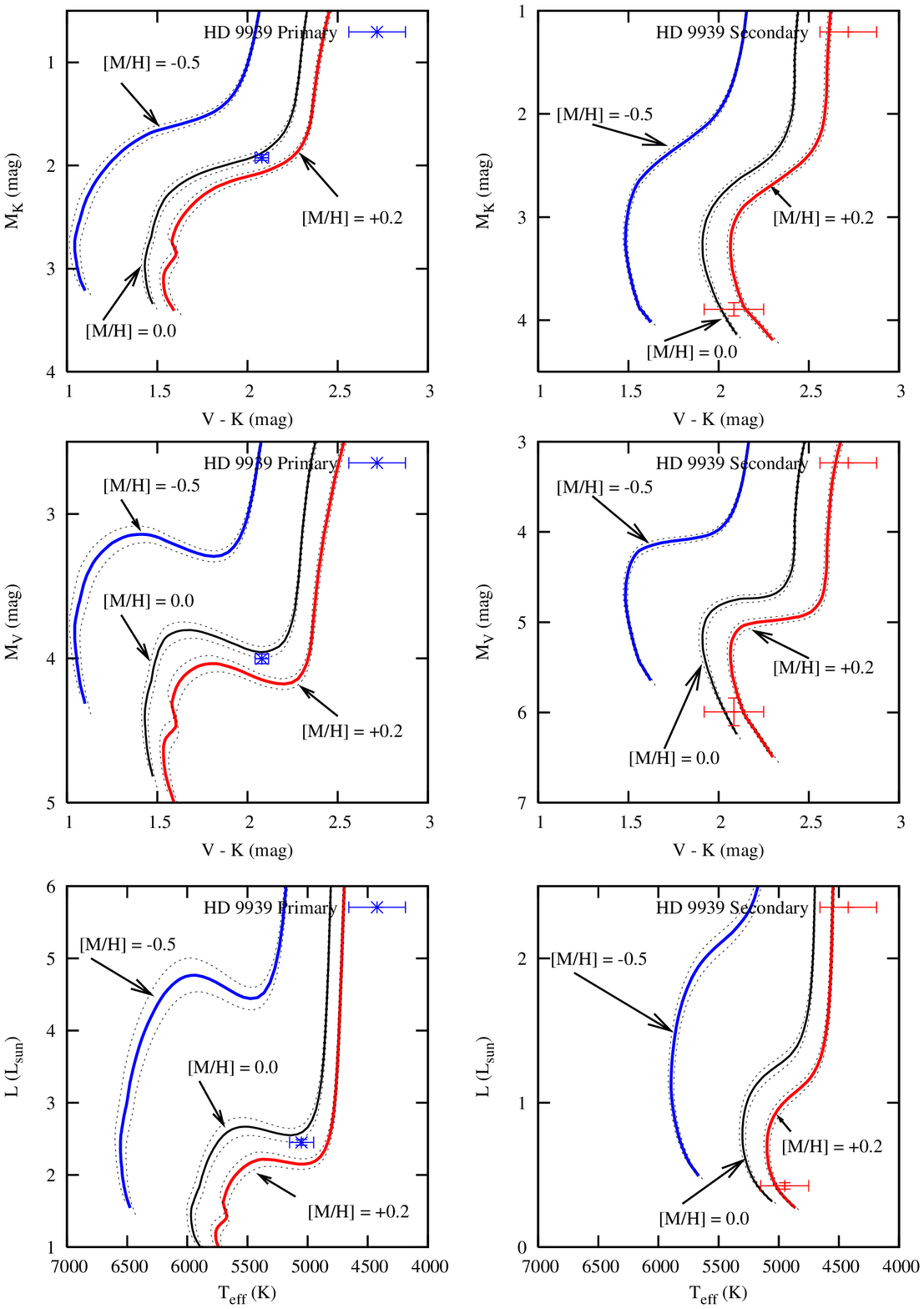}
%%\plottwo{figures/hd9939.primary.MKVmK.eps}{figures/hd9939.secondary.MKVmK.eps}\\
%%\plottwo{f5a.eps}{f5b.eps}\\
%%\plottwo{figures/hd9939.primary.MVVmK.eps}{figures/hd9939.secondary.MVVmK.eps}\\
%%\plottwo{f5c.eps}{f5d.eps}\\
%%\plottwo{figures/hd9939.primary.LTeff.eps}{figures/hd9939.secondary.LTeff.eps}\\
%%\plottwo{f5e.eps}{f5f.eps}\\
\caption{Gross HD~9939 Component/Y$^2$ Model Comparisons in
Color-Magnitude Spaces.  Positions of the HD~9939 components (primary
-- left column, secondary -- right column) are given in empirical and
theoretical color-magnitude spaces.  Y$^2$ evolutionary tracks are
given corresponding to the component dynamical mass estimates (with
one-sigma errors, Table~\ref{tab:physics}) for abundance values of
[M/H] = -0.5, 0, and +0.2.  Clearly the estimated component parameters
are better matches for a roughly solar abundance than either sub- or
super-solar abundances.
\label{fig:hd9939ModelComparisons1}}
\end{figure}

Figure \ref{fig:hd9939ModelComparisons1} gives coarse comparisons
between the inferred properties of the HD~9939 components and Y$^2$
evolutionary tracks computed for a range of elemental abundances in
empirical and theoretical color-magnitude spaces.  The left column of
Figure \ref{fig:hd9939ModelComparisons1} shows comparisons between
observed properties for the HD~9939 primary and evolutionary tracks
computed for the primary dynamical mass (with one-sigma errors;
Table~\ref{tab:physics}) and [M/H] abundances of -0.5, 0.0, and +0.2
dex.  Similarly, the right column gives the corresponding comparisons
for the HD~9939 secondary.  Comparisons with the Y$^2$ models
reinforce the conclusion from our spectroscopic analysis
(\S\ref{sec:spectroscopy}); the elemental abundance for the HD~9939
system is not sub-solar as had been thought
\citep[e.g.][]{Carney1994}, but instead is near the solar value.

\begin{figure}[p]
\epsscale{0.85}
\plotone{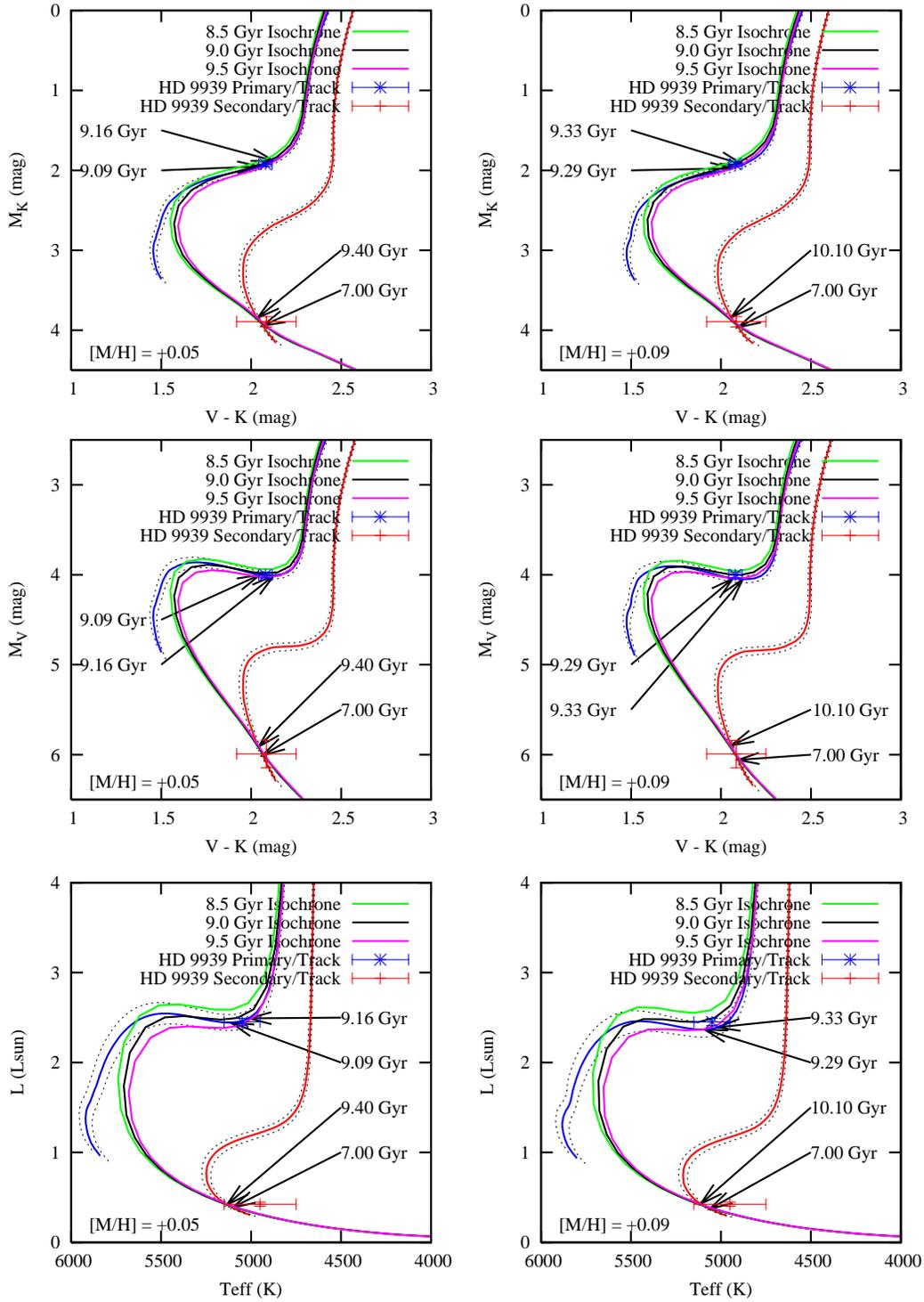}
%%\plottwo{figures/hd9939.isochrone2.MKVmK.eps}{figures/hd9939.isochrone.MKVmK.eps}\\
%%\plottwo{f6a.eps}{f6b.eps}\\
%%\plottwo{figures/hd9939.isochrone2.MVVmK.eps}{figures/hd9939.isochrone.MVVmK.eps}\\
%%\plottwo{f6c.eps}{f6d.eps}\\
%%\plottwo{figures/hd9939.isochrone2.LTeff.eps}{figures/hd9939.isochrone.LTeff.eps}
%%\plottwo{f6e.eps}{f6f.eps}
\caption{Detailed HD~9939 Component/Y$^2$ Model Comparisons in
Observational and Theoretical Color-Magnitude Spaces.  Best-fit Y$^2$
evolutionary tracks and isochrones are given for abundances with
([M/H] = +0.05, left column) and without ([M/H] = +0.09, right column)
inclusion of theory-space parameters in the comparisons.  The range of
track-implied ages for the individual components are marked along the
tracks.  The primary luminosity estimate favors the [M/H] = +0.05/9.1
Gyr hypothesis.
\label{fig:hd9939ModelComparisons2}}
\end{figure}

With the realization that the HD~9939 abundance is approximately
solar, we have used the Y$^2$ models to evaluate the best match in
elemental abundance for the inferred component parameters.  Figure
\ref{fig:hd9939ModelComparisons2} shows the results of that process.
The left column of Figure \ref{fig:hd9939ModelComparisons2} shows the
results of an abundance optimization using all the depicted component
parameters ($V-K$, M$_K$, M$_V$, T$_{eff}$, L).  We find excellent
agreement for all these parameters with an abundance of [M/H] = +0.05.
However if we ignore the inferred T$_{eff}$ and luminosity, we find a
slightly better match for the remaining component parameters at an
[M/H] = +0.09 (Figure \ref{fig:hd9939ModelComparisons2}, right
column).  The Y$^2$ models suggest an elemental abundance for HD~9939
of [M/H] = 0.05 $\pm$ 0.05 dex.

Figure \ref{fig:hd9939ModelComparisons2} indicates that the HD~9939
secondary is on the main sequence, while the HD~9939 primary is
apparently in the Hertzsprung gap, a period of rapid evolution between
the end of the main sequence and the bottom of the red giant branch.
While it is a priori unlikely to find a star in this location in the
HR-diagram (a star with the mass and abundance of the HD~9939 primary
spends only approximately 450 Myrs in this phase, roughly 5\% of the
present system age), there is no apparent alternative to this
conclusion.  This makes the HD~9939 primary a potentially important
test of stellar models.

Finally, Figure \ref{fig:hd9939ModelComparisons2} indicates times
along the evolutionary tracks consistent with the one-sigma excursions
in the measured component parameters.  As the HD~9939 secondary's
radiative properties are relatively poorly measured and its position
on the main sequence (where evolution is relatively slow), a large
range of ages (roughly 7 -- 10 Gyrs) is consistent with the inferred
parameters.  However, the better determined radiative properties of
the HD~9939 primary, coupled with the star's rapid evolution in the
Hertsprung gap results in a smaller viable age range.  For instance,
at a fixed [M/H] of +0.05, the range of ages implied by the Y$^2$
track is only 9.09 -- 9.16 Gyrs (the corresponding range for [M/H] of
+0.09 is 9.29 -- 9.33 Gyrs).  As the system abundance is uncertain at
approximately 0.05 dex, we estimate the HD~9939 primary age at 9.12
$\pm$ 0.25 Gyrs (making no allowance for possible systematic errors in
the Y$^2$ models.)  This estimate is completely consistent with the
large secondary range, so we adopt the primary age estimate for the
HD~9939 system.  Our system age estimate should be more reliable than
the 4.6 Gyr value given by \citet{Wright2004} based on measurements of
chromospheric Ca II H \& K activity; the \citet{Wright2004} estimate
for HD~9939 ignored both the binarity of the system and the subgiant
nature of the primary\footnote{Wright~2006, private communication}.

\section{Discussion}
\label{sec:discussion}

The consensus from our spectroscopic data and component parameter
comparisons with the Y$^2$ models makes it clear that the abundance of
HD~9939 is approximately solar.  Figure
\ref{fig:hd9939ModelComparisons2} shows that the Y$^2$ models do an
excellent job of predicting the parameters for the HD~9939 components,
and suggest that the elemental abundance for HD~9939 is [M/H] = +0.05
$\pm$ 0.05.  These model comparisons further make it clear that the
HD~9939 primary is currently traversing the Hertzsprung gap, and as
such it is currently in a rapid phase of its evolution.  Our
determinations of dynamical mass, luminosity, and abundance for the
primary make it an important empirical constraint for stars in this
phase of evolution.  Earlier provisional assessments of sub-solar
abundance for HD~9939 were apparently biased by the unmodeled subgiant
nature of the primary\footnote{Carney~2005, private communication}.

The combination of the HD~9939 primary's evolutionary state and the
precision of our dynamical mass and radiometric parameter
determinations allow us to estimate the HD~9939 age with surprising
precision at 9.12 $\pm$ 0.25 Gyrs.  This age is qualitatively
consistent with the high space motion of the HD~9939 system at 82.51
$\pm$ 0.91 km s$^{-1}$ with respect to the LSR
\citep[U = 80.32 $\pm$ 0.41
km s$^{-1}$, V = -18.13 $\pm$ 0.60 km s$^{-1}$, W = -5.34 $\pm$ 0.55
km s$^{-1}$, using conventions from][]{Johnson1987}.

Kinematically it is unclear whether HD~9939 is a member of the
galactic thin disk or galactic thick disk; its position in a kinematic
Toomre diagram \citep[e.g.][]{Fuhrmann2002,Venn2004} is shared
between these two populations \citep[however][discuss pitfalls of such
kinematic classification]{Nordstrom2004}.  Based on the velocity
ellipsoid models of \citet{Venn2004} we compute a relative thick/thin
disk probability of 65\%/35\%.  If HD~9939 is a thick disk member then
it would be among the youngest and most metal-rich members of that
population.  Independent of population interpretation, HD~9939 is
notable in abundance for a star its age.  For instance, the
age-metallicity studies of \citet{Rocha-Pinto2000} found no stars the
age and metallicity of HD~9939 in their sample.  However, prior
\citep[notably][]{Edvardsson1993} and subsequent
\citep[e.g.][]{Feltzing2001,Nordstrom2004} studies have found examples
of stars with similar ages and metallicities as HD~9939.  Narrowly the
parameters of the HD~9939 system clearly support the tenets that old,
metal-rich stars do exist, and there is large metallicity dispersion
at all stellar ages.

Old, metal-rich stars suggest that portions of the galactic disk
reached significant metallicity early in its evolution
\cite[e.g.~see][]{Feltzing2001}.  HD~9939's well-determined age and
abundance place potentially interesting constraints on such formation
scenarios.  Based on an epicycle approximation to HD~9939's galactic
orbit \citep{BT87,Makarov2004}, the velocity of HD~9939 implies minimum
and maximum galactic radii of roughly 5.0 and 9.5 kpc respectively,
and a mean galactic radius of 7.3 kpc (with R$_0 \sim$ 8.0 kpc).
Taking the mean radius as the likely formation radius for HD~9939
\citep[following argument from][]{Rocha-Pinto2004}, that roughly
solar-abundance stars were forming 9 Gyrs ago 7.3 kpc from in the
galactic center is remarkable.

\acknowledgements 

The authors thank the anonomous referee for thoughtful contributions
to the presentation of the results presented here.  We also thank
Bruce Carney and Jason Wright for candid and cordial discussions on
various technical aspects of their prior work on the HD~9993 system.
Work done with the Palomar Testbed Interferometer was performed at the
Michelson Science Center, California Institute of Technology under
contract with the National Aeronautics and Space Administration.
Interferometer data were obtained at Palomar Observatory using the
NASA Palomar Testbed Interferometer, supported by NASA contracts to
the Jet Propulsion Laboratory.  Science operations with PTI are
conducted through the efforts of the PTI Collaboration
(http://huey.jpl.nasa.gov/palomar/ptimembers.html), and we acknowledge
the invaluable contributions of our PTI colleagues.  We particularly
thank Kevin Rykoski for his professional operation of PTI.  We thank
Joe Caruso, Bob Davis, Robert Stefanik, and Joe Zajac for obtaining
many of the spectroscopic observations used here.  Thanks also to
Valeri Makarov for valuable discussions concerning galactic orbits.
GT acknowledges partial support from NASA's MASSIF SIM Key Project
(BLF57-04) and NSF grant AST-0406183.  This research has made use of
services of the MSC at the California Institute of Technology; the
SIMBAD database, operated at CDS, Strasbourg, France; of NASA's
Astrophysics Data System Abstract Service; and of data products from
the Two Micron All Sky Survey, which is a joint project of the
University of Massachusetts and the Infrared Processing and Analysis
Center, funded by NASA and the National Science Foundation.

\end{document}